# Comment on "Compton Scattering Spectrum for Si and Ge Systems"
Chen-Kai Qiao *et al*. DOI 10.1088/1361-6471/ab5e35

Compton scattering of photons from a free electron at rest, is well described by the Klein-Nishina[1] theory which predicts a single monoenergetic energy, for the scattered photon, dependent on the scattering angle of the photon:

$$\omega_f = \frac{\omega_i}{1 + \omega_i(1 - cos\theta_{ph})/m_e c^2}$$

The scattering cross section is obtained from two simple Feynman diagrams (second order QED), for example see the book of Bjorken and Drell[2]. This is an unrealistic situation because free electrons at rest do not exists, even in hot plasmas where they are unbound, they are moving. In general electrons are bound in atoms. It can be argued that at incoming energies much larger that the bound energy they are well approximated by free electrons but 1) they are still moving and 2) at small momentum transfers where the transferred energy is small the binding plays an important role. The dependence of the Compton cross section on the momentum transfer is corrected by the incoherent scattering factor (ICF) - *S(q,Z)*[3] (*q* is the absolute value of the momentum transfer) – calculated with atomic wavefunctions. The final energy is still monoenergetic but the cross section (for an atom) is changed to:

$$\left.\frac{d\sigma}{d\Omega_{ph}}\right) = \left.\frac{d\sigma}{d\Omega_{ph}}\right)_{KN} S(q,Z)$$

*S(q,Z)* approaches *Z* at large momentum transfers and goes to zero when *q* approaches zero. This is the basic formulation used by the well known NIST code XCOM[4] and by the tables of the incoherent cross sections available at the NIST site (Compton scattering goes also by the name Incoherent scattering in distinction to the elastic scattering which is a coherent process involving sums of amplitudes, see Ref. 5).

The equation above is still unsatisfactory compared with the experiment. Fig. 1 (taken from[6]) presents a Compton scattering measurement (an energy spectrum) from LiH. The incoming energy was 59.9 keV (from a [241]Am source) and the scattering angle about 160°. The right peak at this energy is the elastic (Rayleigh) scattering. It is somewhat broadened due to the final resolution of the detector (a pure Ge planar detector). The big peak on the left is the Compton scattered peak. It is much more broadened, compared with the elastic peak. This additional broadening is a *Doppler Broadening*, it comes from the movement of the electrons. That means that the scattered photon returns at different energies not just the monochromatic energy predicted by the Klein-Nishina and, hence, the cross section is a

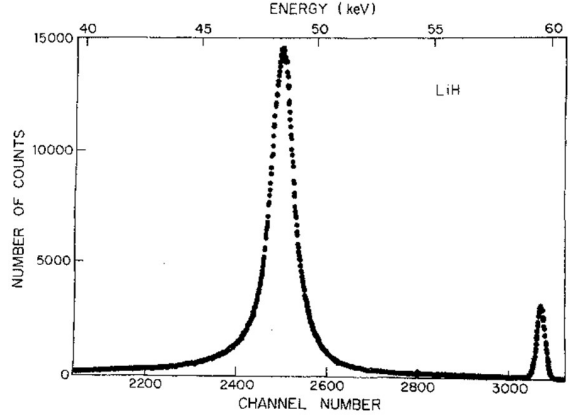

double differential cross section (DDCS) depending both on the scattering angle and the final photon energy:

$$\frac{d^2\sigma}{d\Omega_{ph}d\omega_f}$$

Eisenberger and Platzman[7], based on previous works by DuMond[8], gave a formulation for the above cross section based on the Impulse Approximation (i.e., the interaction photon-electron is very fast, hence, the electron feels the same potential before and after the scattering). In this formulation the DDCS is given as a prefactor (obtained essentially from the kinematics) times a quantity called Compton Profile (CP) – $J(p_z)$, which is the projection of the electron momentum distribution on a z-axis, taken as the axis of momentum transfer:

$$J(p_z) = \iint n(\boldsymbol{p})dp_x dp_y,$$

where *n(p)* is the electrons momentum density. The CP encapsulates the physics of the electrons in the target; it depends on the momentum component $p_z$ hence on energy, explaining, at least qualitatively the energy dependence in Fig. 1.

The CP and the prefactor given by Eisenberger and Platzman were obtained non-relativistically (NRIA). Later, Ribberfors[9] obtained a relativistic formulation (RIA) employing some additional assumptions.

Recently Chen-Kai Qiao et al.[10] employed state-of-the-art Dirac-Fock relativistic wavefunctions to calculate the CP. They integrated the DDCS (in RIA) over $d\Omega_{ph}$ in order to obtain the differential cross section $d\sigma/dT$ where $T=\omega_i - \omega_f$ is the energy transfer. They presented the results in their Fig. 1 for Ge, at two energies, 661.7 keV and 122.1 keV. A comparison is done with a calculation based on the incoherent scattering function. The implication is that this is a very new result of some (or great) importance for dark matter discovery experiments.

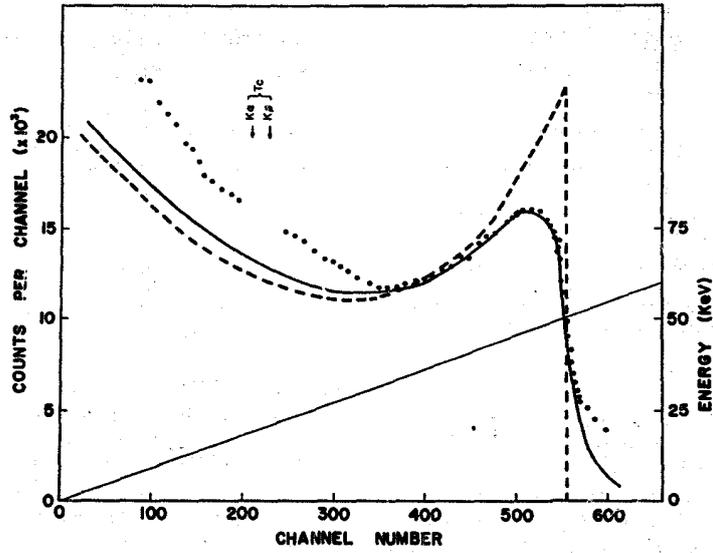

Fig. 2. The Compton edge of a $^{99m}$Tc source. The dashed line is obtained by scattering of free electrons, the solid line by bound electrons (see text).

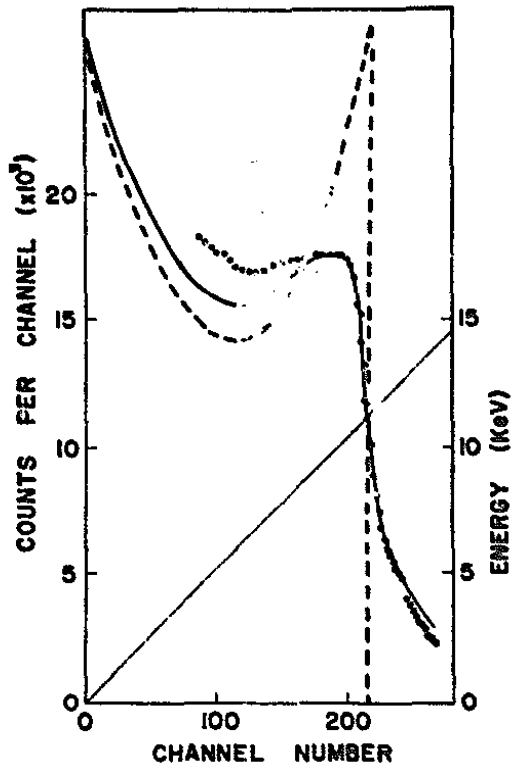

Fig. 1. The Compton edge of a $^{241}$Am source. The dashed line is obtained by scattering of free electrons, the solid line by bound electrons (see text).

I want to point out that this problem was explored almost 50 years ago[11] in relation with using the Compton edge for calibrating plastic detectors. At the time the Compton edge was assumed to be sharply defined as implied by the KN+ICF formulation. By measuring directly, the energy spectrum of a monoenergetic source one obtains the quantity $d\sigma/dT$. For the present purposes of putting in evidence the Doppler Broadening, the intrinsic resolution of the detector has to be small. Therefore, such a measurement has to be done with thin Silicon or Germanium detectors. We not only calculated the detector response but *measured it* and compared between the two. Our results for a $^{99m}$Tc (142.7 keV) and for a $^{241}$Am (59.6 keV) sources, measured with a Si(Li) detector, are shown above. The dots are the measured points, the full line the NRIA calculation, the dashed line the KN+ICF calculation. The diagonal line is the electron energy calibration, with its scale on the right side.

The maximum energy of the electron in the $^{241}$Am case is less than 15 keV corresponding to $\beta=v/c$ of about 0.24, therefore the non-relativistic approach, in this case, was sound. In the case of $^{99m}$Tc this value is 0.5 and clearly a relativistic approach, which was not available at the time, will be more beneficial. The tip of the Compton edge is very well described in both cases. The discrepancies at lower energies are most probably due to multiple scattering. Methods for evaluating and correcting the MS were developed some time after this paper was published.

In their Fig. 3, Chen-Kai Qiao *et al.* make the case that in the low energy region (< 5 keV for the specific case of Ge at 661.7 keV) the above energy spectrum will contain some small steps as a direct indication of forbidden scattering when the energy transfer is smaller than the binding energy. Jugging by the experimental results of our previous measurements above, even if they are old, it will be a difficult enterprise to actually see these predicted steps at such small energies. The case is that steps where already seen, much easier, not in the energy transfer spectrum but in the DDCS spectrum like one in Fig. 1 above.

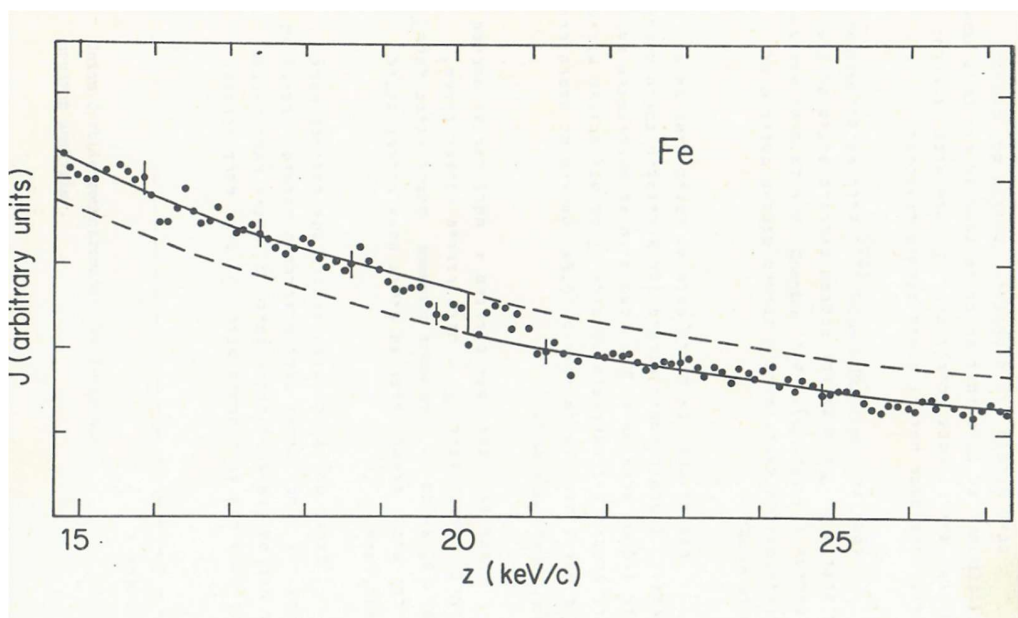

In the above figure (taken from[12]) I present the high momentum part (i.e., the region close to the elastic scattering peak) of the Compton scattered spectrum from Iron. A step is clearly present where the $1s^2$ (K) electrons stop contributing to the scattering due to energy consideration. Other steps are not seen due to two reasons 1) the original measurement was not intended to find steps, the statistics achieved with a small lab $^{241}$Am source being insufficient 2) the electrons in the higher shells are more expanded in the real space and concentrated in a smaller region in the momentum space, hence their contribution at high momentum values is smaller. But, because their binding energy is smaller, their step should appear closer to the elastic scattering peak, i.e., in the high momentum region. Probably, with high fluence synchrotron sources other steps can be put in evidence.

Summary: A major part of the theoretical prediction of Chen-Kai Qiao *et al.* were verified experimentally a long time ago.

Sylvian Kahane – retired from Physics Dept., NRCN, Israel
sylviankahane@gmail.com